\documentclass[]{spie}  
\usepackage[]{graphicx}
\begin{document}
\title{The SXI telescope on board EXIST: scientific performances}


\author{L. Natalucci\supit{a}, A. Bazzano\supit{a}, 
S. Campana\supit{b}, P. Caraveo\supit{c}, R. Della Ceca\supit{b}, 
J.E. Grindlay\supit{d}, F. Panessa\supit{a}, G. Pareschi\supit{b}, 
B. Ramsey\supit{e}, G. Tagliaferri\supit{b}, P. Ubertini\supit{a},  
G. Villa\supit{c}
\skiplinehalf
\supit{a}INAF-Istituto di Astrofisica Spaziale e Fisica Cosmica, 
Sezione di Roma, via Fosso del Cavaliere 100, I-00133 Roma, 
Italy; \\
\supit{b}INAF-Osservatorio di Brera, 
via Bianchi 46, I-23807 Merate, Italy; \\
\supit{c}INAF-Istituto di Astrofisica Spaziale e Fisica Cosmica,
Sezione di Milano, Via E. Bassini 15, I-20133 Milano, Italy \\
\supit{d}Harvard-Smithsonian CfA, 
60 Garden Street, Cambridge MA 02138, USA \\
\supit{e}NASA Marshall Space Flight Center, VP62, Huntsville, 
AL 35812, USA \\
}

\authorinfo{Further author information: (Send correspondence to L.Natalucci)\\L.Natalucci: E-mail: Lorenzo.Natalucci@iasf-roma.inaf.it, Telephone: +39 06 4993 4461}

 
  \maketitle

\begin{abstract}

The SXI telescope is one of the three instruments on board EXIST, 
a multiwavelength observatory in charge
of performing a global survey of the sky in hard X-rays searching
for Supermassive Black Holes. One of the primary objectives of EXIST
is also to study with unprecedented sensitivity the most unknown high
energy sources in the Universe, like high redshift GRBs, which will be
pointed promptly by the Spacecraft by autonomous trigger 
based on hard X-ray localization on board. The recent addition of a 
soft X-ray telescope to the EXIST 
payload complement, with an effective area of ~950 cm$^2$ in the energy 
band 0.2-3 keV and extended response up to 10 keV will allow to make
broadband studies from 0.1 to 600 keV.  
In particular, investigations of the spectra components and states of AGNs
and monitoring of variability of sources, study of the prompt and afterglow 
emission of GRBs since the early phases, which will help to constrain the 
emission models and finally, help the identification of sources in the EXIST 
hard X-ray survey and the characterization of the transient events detected.
SXI will also perform surveys: a scanning survey with sky coverage 
$\sim2$~$\pi$ 
and limiting flux of $\sim5\times10^{-14}$ cgs plus other serendipitous.
We give an overview of the SXI scientific performance and also describe 
the status of its design emphasizing how it has been derived by the scientific 
requirements.       
\end{abstract}

\keywords{X-ray telescopes, EXIST, X-ray Instrumentation}

\section{INTRODUCTION} \label{sect:sections}

The potential of surveys in high energy astronomy is being fully demonstrated by the sky 
observations performed at X- and soft $\Gamma$-ray wavelengths during the last decade.
In particular XMM Newton$^1$ 
with its deep sensitivity in the soft X-ray band, INTEGRAL$^2$ 
and Swift$^3$ 
up to $>100$ keV have started to reveal 
the demographics of sources in the near Universe by studying populations of objects. 
Still lacking today are both a comprehensive picture  
of transient phenomena (other than GRBs) spanning a broad range of durations and a deep
study of the supermassive objects and their host galaxies against Universe age. 
Studies of transients require 
fast repointing and improved sensitivity for proper investigations of their prompt
emission. The fast follow-up capability of Swift allows to probe the farthest regions of the Universe by the study of Gamma-Ray Bursts (GRB) and also to discover new populations of transients in our Galaxy. GRBs are today the only beacons that can be used to observe
the Universe at $z>3-4$, as demonstrated by Swift e.g. with the latest detection of a GRB 
at z=8. Swift has opened a new window on the far Universe with
GRBs, but many more objects are needed including AGNs. 

Swift also carries an X-ray telescope which is an unvaluable tool to 
observe the GRB afterglows (and Swift has demonstrated that almost all the GRBs have X-ray
afterglows).  The Swift/XRT telescope$^4$
is also a powerful tool for localization
of the sources seen by Swift/BAT and INTEGRAL/IBIS$^5$
 at hard X-ray energies. 
In the case of IBIS, of the 167 new sources in the 3rd IBIS catalog$^6$ 
129
have been identified through optical and NIR spectroscopy. Of these, more than 60\% 
have been localized by XRT with typical accuracy $<2-3$ arcsec. Most of the reasons for this
success is the possibility for XRT to perform many, relatively short observations. 

The concept of operability of Swift is also adopted in the design of the Energetic X-ray Imaging Survey Telescope (EXIST) mission$^7$.
EXIST is a proposed hard X-ray imaging all-sky deep survey mission and was recommended by 2001 Report of the Decadal Survey. It is a strong candidate to be the Black Hole Finder Probe, one of the three "Einstein Probes" in the Beyond Einstein Program, now proposed for the Astro2010 Decadal Survey. 
EXIST will be launched in a Low Earth Orbit (LEO) and its primary instrument is the 
High Energy Telescope (HET), a wide 
field coded aperture instrument covering the 5-600 keV energy band and imaging sources 
in a $\sim70\times90$deg$^2$ field of view with 2~arcmin resolution and better than 20~arcsec positions$^8$. 
The energy band of HET overlaps with the soft X-ray range 
covered by the proposed Soft X-ray Imager (SXI), 0.1-10 keV with an effective area of
950cm$^2$ at 1.5 keV and 3.5m focal length. At longer wavelengths, 
the IRT$^9$ is an optical-IR aperture telescope covering the $0.3-2.2$ micron 
range with variable spectral resolution and high sensitivity (AB=24 in 100s).
The IRT pixel 
size is 0.15~arcsec and its Field-of-View in Imaging mode is 16~arcmin$^2$. So EXIST is
a real multiwavelength observatory for observations of GRBs and Supermassive Black Holes
(SMXB) as well as of many other types of transients and high energy sources. 

In the following we will address some of the topics that will be contributed by the
SXI telescope in the study of the high energy sources seen by HET and its capabilities 
for sky survey at soft X-ray wavelengths. Section 2 contains a brief description of the 
SXI telescope and its effective area, Section 3 and 4 will describe some of the science 
cases relevant to SXI.  

   \begin{figure}
   \begin{center}
   \begin{tabular}{c}
   \includegraphics[height=11cm]{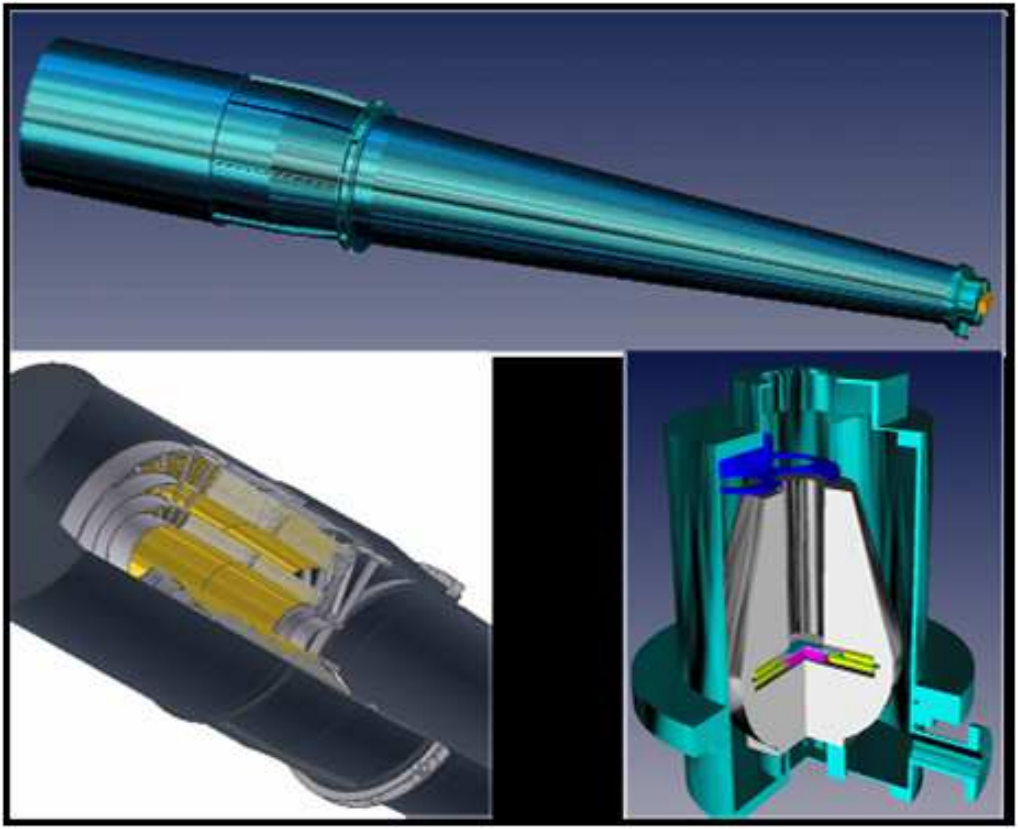}
   \end{tabular}
   \end{center}
   \caption[example]
   { \label{fig:sxi_telescope}
Design overview of the SXI telescope. The instrument (top) has a mirror focal
length of 3.5m and the overall length is 4.5m, 70 cm
max width. The primary mirror system and the camera (bottom) are also shown. 
}
   \end{figure}
 
\section{THE SXI DESIGN} 
 
The proposed design is based on a Wolter type-I telescope consisting of a main mirror assembly with 26 nested cells and a focal plane camera with CCD detector. The focal plane distance
is 3.5m and the max.diameter of the mirrors
is 60cm (giving 70cm on the telescope outer envelope). The telescope structure, shown
in Fig 1 (top) is built around an I/F flange in titanium which is the unique interface to the satellite optical bench. The mirror system and camera are also shown in the same figure
(bottom).

The structure consists of a forward tube holding the mirrors and a rear conical tube
holding the camera, with a single spider on which the mirrors are mounted. Below the spider 
there will be an electron diverter and on top of the mirror system a pre-collimator will
reduce the intensity of stray light. The camera design is very similar to 
the XRT and XMM-EPIC design. A CCD sensor and its Proximity Electronics are "suspended" in an Al shield and an active cooling system will ensure the optimal temperature for CCD operation. The sensor
will be required to have a frame readout between 5 and 10ms (see Section 3). The camera 
contains other subsystems like the Vacuum Chamber, Filter Wheel and 
Vacuum Door. More
information on the technical design of SXI can be found in a related 
paper$^{10}$. 

   \begin{figure}
   \begin{center}
   \begin{tabular}{c}
   \includegraphics[height=8cm,angle=0]{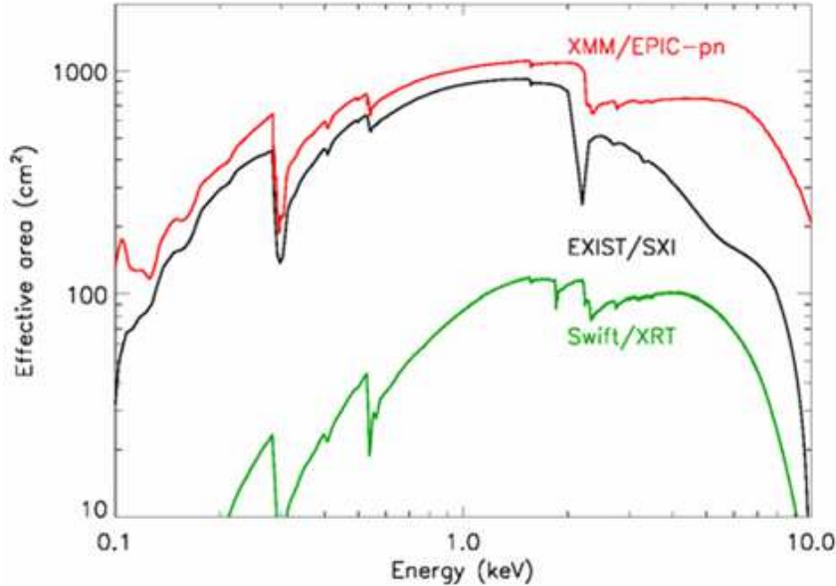}
   \end{tabular}
   \end{center}
   \caption[example]
   { \label{fig:effective_area}
SXI effective area baseline configuration,
compared to that of a XMM single telescope and of the
XRT telescope onboard Swift. }
   \end{figure}

In Figure 2 is shown the effective area of the telescope, obtained convolving the
effective area of the mirrors with a suitable filter and with the response of the
pn-CCD similar to XMM. The effective area is similar to the one of an XMM module 
for $E<2$ keV, losing efficiency at higher energies due to lower focal length. The
parameters of the baseline design of SXI are reported in Table 1.

\begin{table}
\caption{\label{tab:fov} Design parameters of the SXI telescope.}
\begin{tabular}
{   l   |  l   }             
\hline
Parameter      &   Baseline    \\ 
\hline
Mirror                    &    26 shells             \\
Angular Resolution        &    20~arcsec at 1 keV     \\
Energy Range              &    0.1-10 keV            \\
Diameter of Mirror        &    60cm    \\
Focal Lenght              &    3.5m    \\
Detector Type             &    {\em pn}-CCD  \\
Detector Size             &    3x3 cm$^2$   \\
FOV                       &    30x30 arcmin$^2$  \\
Energy Resolution         &    130eV at 6 keV   \\
Readout Speed             &    5-10 ms    \\
Effective Area            &    950 cm$^2$ at 1.5 keV, $>100$ cm$^2$ at 8 keV \\
Sensitivity ($10^{4}s$)    &    $2\times10^{-15}$ cgs   \\
\hline
\end{tabular}
\end{table}

   \begin{figure}
   \begin{center}
   \begin{tabular}{c}
   \includegraphics[height=10cm]{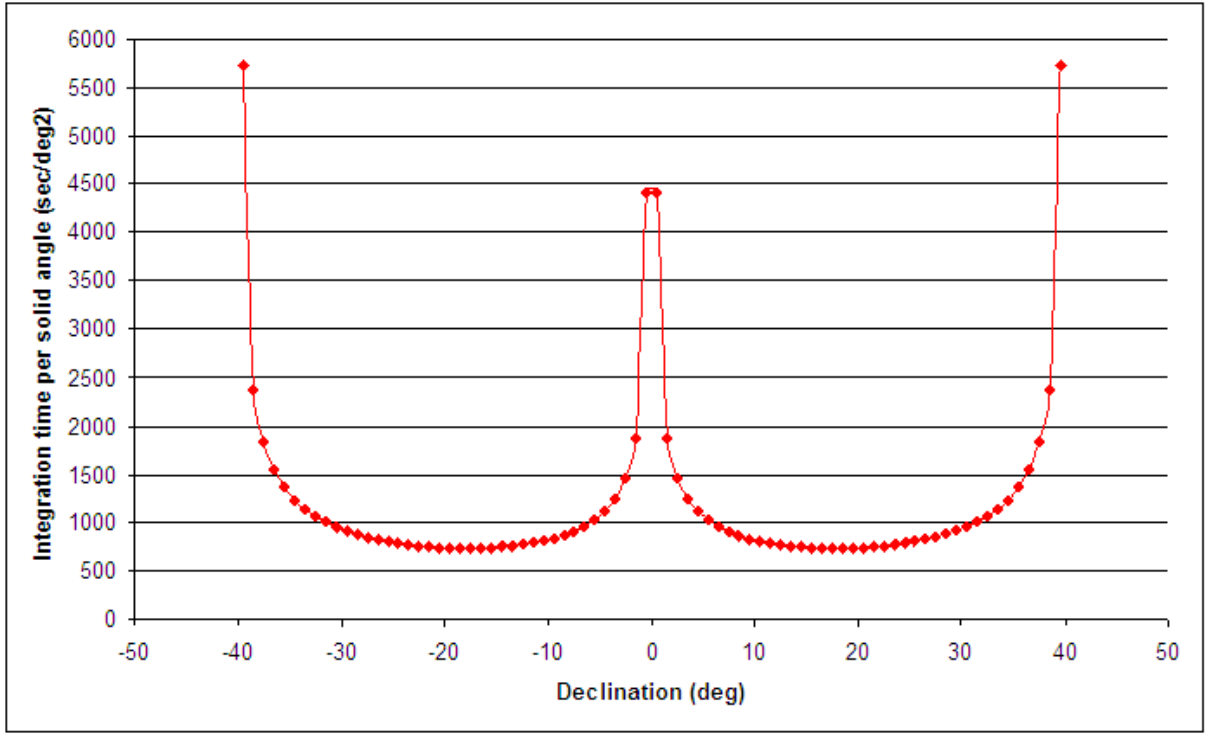}
   \end{tabular}
   \end{center}
   \caption[example]
   { \label{fig:sxi_coverage}
Sky coverage/square degree, for a narrow field instrument during one year of HET 
scanning survey. The coverage is seen as a function of declination. A large fraction 
of the sky, comprised between -40 and 40 degrees declination will be amenable to
SXI observation. The average time for each source in a SXI field will be $\sim200s$
each year. 
}
   \end{figure}

\section{OPERATION OF THE SXI TELESCOPE} \label{sect:sections}

In analogy with Swift/XRT, SXI will operate with different modes, the count rate threshold
of which basically depends on the performance of its detector. A Photon Counting (PC) mode will be used to observe sources up to $\sim50-100$ mCrabs, whereas more intense sources will be monitored by means of other modes allowing for pileup. The main difference with respect to 
the XRT telescope
is the possibility of operation during continuous scans. During the first two years EXIST
will be programmed for a full sky survey of HET. With its large FOV the HET is capable of 
a full sky coverage each two orbits, by performing a continuous scan while pointing at the
zenith. The last three years will be instead dedicated to inertial pointing observations,
mainly of the sources detected by the HET in the first two year's survey.

\subsection{The SXI Scanning Survey} 

During HET scanning, the spacecraft is zenith-pointed with an offset of 
30 degrees, towards the north and towards the south on
alternate orbits. In the meanwhile, SXI will record events and the fast readout of its
detector will allow to locate their direction within a few arcsec.  
The coverage of the SXI scanning survey is shown in Figure 3. In the plot is shown 
the sky coverage (calculated per square degree) of a narrow field instrument co-aligned 
with HET during one year of zenith scanning. The resulting average exposure time for a
source being of the order of 200s/year, it is estimated that SXI will cover half of the sky 
with a limiting sensitivity of $\sim5\times10^{-14}$ erg~cm$^{-2}$~s$^{-1}$ in two years. 
The limiting sensitivity of the survey allows to investigate a high number of sources
(a few hundred thousand). 

As during the scanning survey the IRT is not expected to be active, SXI will be able to
improve localization of many faint HET sources detected during the first two years:
from $\sim20$~arcsec to a few arcsec. 
 
   \begin{figure}
   \begin{center}
   \begin{tabular}{c}
   \includegraphics[height=8cm]{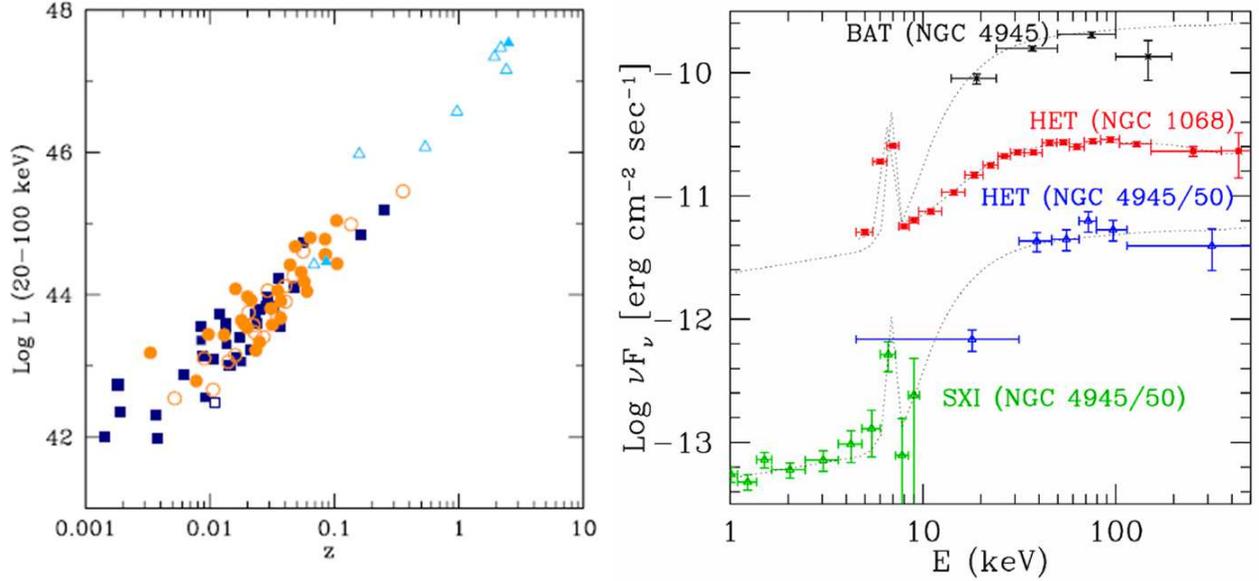}
   \end{tabular}
   \end{center}
   \caption[example]
   { \label{fig:agn_spectrum}
{\em Left}: the redshift distribution of a sample of 88 AGNs seen by INTEGRAL/IBIS: 
type I objects are represented by circles, type II by squares and blazars by triangles.
From Malizia et al. 2009. {\em Right:} simulation of spectra of known 
Compton thick AGNs by SXI and HET on board EXIST. HET ones are simulated survey spectra
while SXI data are assumed for a $10^{4}$s snapshot.
}
   \end{figure}

\subsection{Pointing and Serendipitous Surveys} 

During the second phase of the EXIST mission, lasting 3 years it is expected to point
at individual sources to acquire 
broadband images and spectra of the HET survey objects. This will allow for
simultaneous broadband observations covering optical/NIR, soft X-ray to soft 
$\gamma$-ray
bands. The SXI telescope as well as the IRT will measure source intensities and spectra
for the hard X-ray sources observed by HET down to a sensitivity limit of 
$\sim0.1$ mCrab in the hard X-ray band. 

During observations of individual targets, SXI will be able to detect serendipitous 
sources in its field of view, at least if the initial target flux is not bright enough
to allow operation in Photon Counting mode. If the typical observation time for this
pointing phase is assumed to be $\sim10^4$s, then assuming $\sim2000$ observations 
each year it will be possible to 
cover $\sim1500$~cm$^2$ during the 3 years of the 
inertial pointing phase, down to the sensitivity limit of 
$\sim2\times10^{-15}$~erg~cm$^{-2}$~s$^{-1}$. For comparison,
the most populated catalog ever made at X-ray wavelengths is that of XMM-Newton, with
191,870 sources and a sky coverage of 360 square degrees$^{11}$. 
Another serendipitous 
survey will be based on the detection of serendipitous sources for GRB 
repointing.

   \begin{figure}
   \begin{center}
   \begin{tabular}{c}
   \includegraphics[height=7cm]{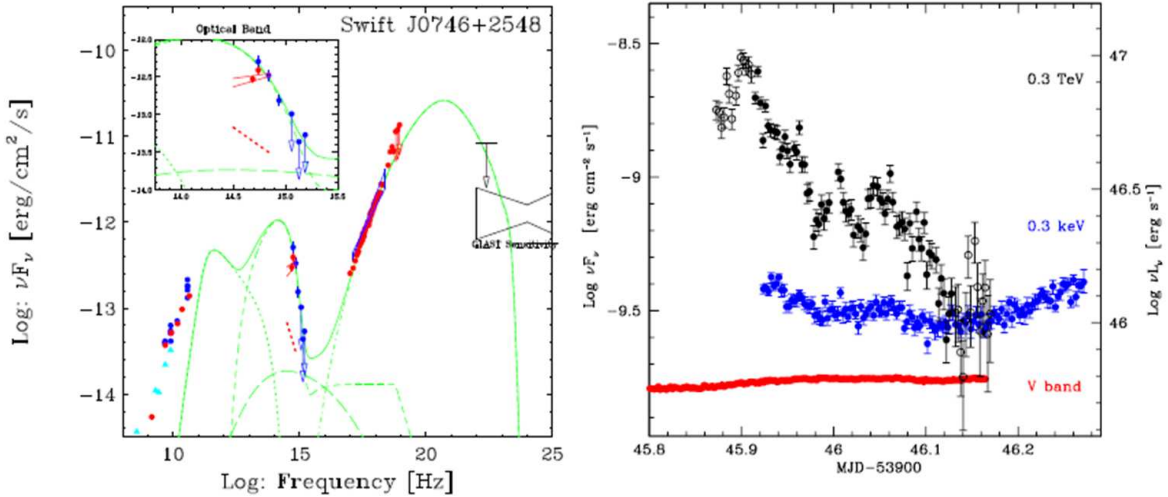}
   \end{tabular}
   \end{center}
   \caption[example]
   { \label{fig:blazars}
{\em Left:} Blazar SED of SWIFT J0746+2548 with X-ray data from Suzaku and Swift/XRT (from
Watanabe et al. 2009).
{\em Right}: Simultaneous light curves of the high peaked BL~Lac source PKS 2155-304 by HESS, Chandra and Bronberg Opt. Observatory (from  Costamante 2009).
}
\end{figure}

\section{STUDIES OF INDIVIDUAL POINT SOURCES} \label{sect:sections}

EXIST being a powerful multiwavelength observatory, it will allow to 
obtain 0.1-500 keV broadband spectra for $\sim500$ or even more AGN with extension to the optical/NIR band. In Figure 4 (left panel) is shown the luminosity/redshift 
distribution of the AGNs detected by IBIS, corresponding to a survey with a limiting
sensitivity of $\sim1$~mCrab in the 20-100 keV band$^{12}$. The most distant objects 
detected so far are the blazars, while type I and II objects, including absorbed 
and Compton thick sources are located in the region $z<0.2$.

\subsection{Broadband spectra for AGN studies}

EXIST can study in much more detail the above objects in both number and quality of broadband
spectra. Example spectra of individual objects observations are shown 
in Figure 4 (right panel) for two Compton thick AGNs. Compton thick (CT) sources with 
column densities ${N}_{h}$ $>10^{24}$ are difficult to observe in X-rays and the known
objects are detected at relatively low redshifts$^{13}$, i.e. $z<0.05$. Nevertheless 
CT AGNs are important to study due to their still unknown number and contribution to the 
cosmic X-ray background$^{12,14}$.  

In particular, we show a simulation of SXI+HET observation for the source 
NGC 4945, having a BAT measured flux of $\sim3\times10^{-10}$~erg~cm$^{-2}$~s$^{-1}$ 
(15-200 keV). A source with same spectrum and flux divided by 50 can be detected by
SXI and its spectrum resolved into components.    

\subsection{Studies of variability and SED of Blazars}

In the study of blazars are recognized as very important the characterization of their
spectral energy density (SED) and the monitoring of their variability. Blazars have
relativistic jets pointing at us. The SED of blazars
from radio to $\gamma$-rays is characterized by two broad peaks ascribed mainly to 
synchrotron and Inverse Compton emission (see Figure~5, left panel, for an example SED$^{15}$). Depending on the 
location of the two peaks, 
they can be classified as belonging to the Low-, Intermediate- of High-Energy peaked 
class. The X-ray band can often fall into either the 
high energy tail of the synchrotron peak or the low energy tail of the IC peak, with very different spectral slopes. For objects with insufficient multiwavelength coverage, 
X-ray spectra from 0.1 to 100 keV as could be provided by SXI and HET can be sufficient to constrain or classify them as belonging to either classes. Furthermore as the SED 
can be variable, X-ray spectra slopes can provide constraints on its shape throughout
very long, continuous time span.

An example of variability studied in multifrequency is 
reported in Figure~5, 
right panel for PKS~2155-304.  For this source the large variability detected 
in $\gamma$-rays, 
but not in optical/X-ray suggests probably two SED emitting zones, 
of which only one is responsible for the flaring emission$^{16}$. 

Therefore the study of the variability of Blazars is a clue to unveil the properties of the
emission region. Multiwavelength coverage with time as could be provided by EXIST, each 
time a blazar will be pointed and fast follow-up of flares will be unvaluable to test 
models.

   \begin{figure}
   \begin{center}
   \begin{tabular}{c}
   \includegraphics[height=7cm]{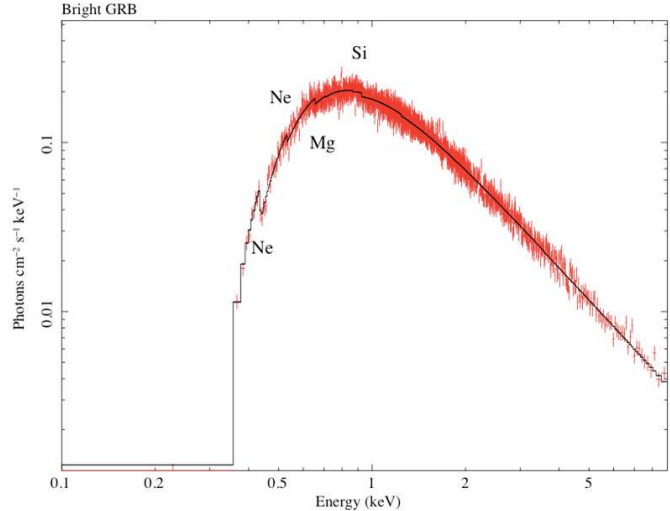}
   \end{tabular}
   \end{center}
   \caption[example]
   { \label{fig:GRB_withedges}
Simulated spectrum of a bright burst spectrum as seen by SXI with absorption edges.
See text for details.
}
   \end{figure}

\subsection{Investigations of GRB early spectra and afterglows}

Swift/XRT observations have raised a lot of new issues about the origin of the emission mechanisms for the
prompt and late (afterglow) emission in GRBs. Since many GRB afterglows could be emitted by an 
ISM environment, a sensitive measurement of the X-ray afterglow and detection of features in the early
afterglow emission could help to unveal the structure of the medium surrounding the central engine. In particular, 
the X-ray emission can be used as a tool to investigate the properties of the circumburst medium which imprints
characteristic absorption edges on the X-ray spectrum
through which element abundances can be computed, and some constraints on the evolution of the progenitor
can be set. 

This "edge diagnostics" has been applied for one of the closest GRBs observed by Swift, GRB 060218, 
leaving to the conclusion that low Z metals have been ejected from the progenitor before collapse$^{17}$.
A good energy resolution as provided by EXIST/SXI could provide the tool to study the absorption edges 
formed in the circumburst environment for a number of GRBs. The estimates of the edges could also serve as
an estimate to GRB redshift. In Figure 6 we show a simulation of a bright GRB (flux: $10^{-9}$ cgs for 1ks) 
obtained on the basis of the model used for GRB 060218$^{17}$. Many of the absorption edges 
from the model are clearly resolved. 
The simulations also show that for a typical faint XRT burst (a factor 10 lower in intensity) it is still possible 
to obtain measurements of the redshift within an error as low as $\sim6$\%.

\section{CONCLUSIONS} \label{sect:sections}

The SXI telescope on board EXIST will take full advantage of the operational strategy
adopted for the mission, mostly based on surveys and fast follow-up of GRB and transients. 
With SXI, EXIST is a real multiwavelength observatory with a sensitive broadband coverage
at high energies: 0.1-600 keV,  with SXI providing an effective area of
950cm$^2$ at 1.5keV. SXI will perform wide area surveys (scanning \& serendipitous)
and sensitive observation of transient events in the X-ray (e.g. GRB 
afterglows, 
AGN flares). It will also help the identification of HET sources detected 
during the survey, study the absorbed (even Compton thick) AGN Universe and zoom on AGN states.

The heritage of the Swift/XRT, XMM-Newton and INTEGRAL allows to conclude that the SXI performance is appropriate to the profile of the EXIST mission as currently designed. 
Main improvements to Swift/XRT are: factor $\sim10$ effective area, fast detector readout allowing operation during scanning survey.

\section*{ACKNOWLEDGMENTS}

The Italian authors acknowledge the support of ASI contract    
I/088/06/0.
 

\section{REFERENCES}

\begin{enumerate}

\item
N. Gehrels, G Chincarini, P. Giommi, K. O. Mason, J. A. Nousek, A. A. Wells, N. E. White, S. D. Barthelmy, D. N. Burrows, L. R. Cominsky, K. C. Hurley, F. E. Marshall, P. Meszaros, P. W. A. Roming, "The Swift Gamma-Ray Burst Mission", 
ApJ 611, p. 1005, 2004

\item
F.Jansen, D.Lumb, B.Altieri, J.Clavel, M.Ehle, C.Erd, C.Gabriel, M.Guainazzi, P.Gondoin, R.Much,
R.Munoz, M.Santos, N.Schartel, D.Texier, G.Vacanti, "XMM-Newton Observatory. I. The spacecraft and operations", 
A\&A 365, p.L1-L6, 2001

\item
C. Winkler, T. J.-L. Courvoisier, G. Di Cocco, N. Gehrels, A. Giménez, S. Grebenev, W. Hermsen, J. M. Mas-Hesse, F. Lebrun, N. Lund, G. G. C. Palumbo, J. Paul, J.-P. Roques, H. Schnopper, V. Schönfelder, R. Sunyaev, B. Teegarden, P. Ubertini, G. Vedrenne and A. J. Dean,
"The INTEGRAL mission",
A\&A 411, L131, 2003

\item
D.N. Burrows, J.E. Hill, J.A. Nousek, J.A. Kennea, A. Wells, J.P. Osborne, A.F. Abbey, 
A. Beardmore, K. Mukerjee, A.D.T. Short, G. Chincarini, S. Campana, O. Citterio, 
A. Moretti, C. Pagani, G. Tagliaferri, P. Giommi, M. Capalbi, F. Tamburelli, 
L. Angelini, G. Cusumano, H.W. Brauninger, W. Burkert and A.D. Hartner,
"The SWIFT X-ray telescope"
Space Sci. Rev., 120, 165, 2005

\item
P. Ubertini, F. Lebrun, G. Di Cocco, A. Bazzano, A. J. Bird, K. Broenstad, A. Goldwurm, 
G. La Rosa, C. Labanti, P. Laurent, I. F. Mirabel, E. M. Quadrini, B. Ramsey, V. Reglero, 
L. Sabau, B. Sacco, R. Staubert, L. Vigroux, M. C. Weisskopf and A. A. Zdziarski,
"IBIS: the Imager on board INTEGRAL",
A\&A 411, L141, 2003
 
\item
A. J. Bird, A. Malizia, A. Bazzano, E. J. Barlow, L. Bassani, A. B. Hill, G. Belanger, 
F. Capitanio, D. J. Clark, A. J. Dean, M. Fiocchi, D. Gotz, F. Lebrun, M. Molina, 
N. Produit, M. Renaud, V. Sguera, J. B. Stephen, R. Terrier, P. Ubertini, R. Walter, 
C. Winkler, J. Zurita,
"The 3rd IBIS/ISGRI soft gamma-ray survey catalog",
ApJ.Suppl, 170, 175, 2007

\item
J.E. Grindlay, the EXIST Team, 
"GRB Probes of the High-Z Universe with EXIST",
AIPC, Vol. 1133, p.18, 2008

\item
J. Hong et al., these Proceedings, 2009

\item
A. Kutirev et al., Proc. SPIE Conf. 7453, 2009 (in press)

\item
G. Tagliaferri et al., Proc. SPIE Conf. 7437, 2009 (in press)

\item
M.G. Watson, A.C. Schroder, D.Fyfe et al., 
"The XMM-Newton serendipitous Survey. V. The Second XMM-Newton Serendipitous 
Source Catalogue",
A\&A 493, 339, 2009

\item
A. Malizia, J. B. Stephen, L. Bassani, A. J. Bird, F. Panessa, P. Ubertini,
"The fraction of Compton-thick sources in an INTEGRAL complete AGN sample"
MNRAS, in press, 2009 (arXiv 0906.5544)

\item
R.della Ceca, P.Severgnini, A.Caccianiga, A.Comastri, R.Gilli, F.Fiore, 
E.Piconcelli, G.Malaguti and C.Vignali,
"Heavily obscured AGN with BeppoSAX, INTEGRAL, Swift, XMM and Chandra: 
prospects for Simbol-X",
Mem. SAIt Vol.79, p.65, 2008

\item
E.Treister, C.M. Urry, S. Virani
"The Space Density of Compton-Thick Active Galactic Nucleus and the X-Ray Background"
ApJ 696, 110, 2009

\item
S.Watanabe, R.Sato, T.Takahashi, J.Kataoka, G.Madejski, M.Sikora, F.Tavecchio, 
R.Sambruna, R.Romani, P.G. Edwards and T.Pursimo,
"Suzaku Observations of the Extreme MeV Blazar SWIFT J0746.3+2548",
ApJ 694, 294, 2009

\item
L.Costamante,
"Blazar properties: an Update from Recent Results",
Proceedings of the Workshop "High-Energy Gamma-rays and Neutrinos from Extra-Galactic Sources", January 13-16, 2009, to be published in Int. J. Mod. Phys. D, 2009
(arXiv 0907.3967)

\item
S. Campana, N. Panagia, D. Lazzati, A. P. Beardmore, G. Cusumano, 
O. Godet, G. Chincarini, S. Covino, M. Della Valle, C. Guidorzi, D. Malesani, 
A. Moretti, R. Perna, P. Romano, and G. Tagliaferri,
"Outliers from the Mainstream: How a Massive Star Can Produce a Gamma-Ray Burst",
ApJ 683, L9, 2008

\end{enumerate}


\end{document}